\magnification=\magstep1
\vsize=22.7truecm
\hsize=15.5truecm
\hoffset=.6truecm
\voffset=.8truecm
\parskip=.2truecm

\font\ti=cmbx10 scaled\magstep1
\font\eightrm=cmr8

\def\br{\hfill\break\noindent}

\pageno=0

%
%
\baselineskip=.5truecm
\footline={\hfill}
\vskip3truecm
\vskip3truecm
\centerline{\ti Duality Transformations in the Ten-Dimensional
Action}
\centerline{\ti  and New Superstring Theories }
\vskip1.2truecm
\centerline{  A. H. Chamseddine}
\vskip.8truecm
\centerline{ Theoretische Physik}
 \centerline{ ETH-H\"onggerberg}
  \centerline{ CH 8093 Z\"urich }
\centerline{ Switzerland} 
\vskip1.2truecm

\centerline{\bf Abstract}
\vskip.5truecm
\noindent
Type II superstring theory with mixed Dirichlet and Neumann boundary
conditions admit antisymmetric tensors with varying degrees in the
spectrum. We show  that there exists a family of dual supergravity
lagrangians to the $N=2$ type IIA action in ten dimensions. The duality
transformations and the resulting actions are constructed explicitely.
\vskip.5truecm

\noindent

\vfill

\eject
\baselineskip=.6truecm
\footline={\hss\eightrm\folio\hss}

\noindent
At present there are many conjectured dualities between 
supestring theories as well as between superstring and super-p-brane
theories in various dimensions [1-2]. As there is no known way of consistently
quantizing the p-branes giving a massless spectrum, such
conjectures could not be tested. Fortunately there has been a recent
development where it is enough to consider open strings with fixed end
points and
Dirichlet boundary conditions in the presence of
closed strings [3]. The Ramond-Ramond sectors (R-R) of the string
Hilbert space contain vertex operators of the form 
$\overline{Q} \Gamma^{[\mu_1} \cdots \Gamma^{\mu_n ]}Q F_{\mu_1
\cdots \mu_n}$ where $F$ is an n-form field strength. There is
no analysis as yet predicting which mixtures of boundary
conditions for the type II superstring with open strings
having Dirichlet boundary conditions, are consistent and 
what the resulting spectrum is. It should be possible to 
make this study, but we shall attempt to answer this question
by studing the duality transformations that could be carried
on the supergravity action of type IIA in ten dimensions.
Such a procedure has proven its effectiveness in the study
of the $N=1$ supergravity action in ten dimensions [4] where it
was shown that an alternative formulation with the six-form
 replacing the two-form is possible [5]. This even led to simplifications
in deriving the coupling to the super Yang-Mills sector [5] and to 
conjecturing the existence of super five-branes [2]. 

The massless spectrum of the type IIA superstring in ten-dimensions
 is easy to state.
In the NS-NS sector we have the graviton the antisymmetric two-form
and the dilaton. The R-R sector contains an abelian vector (a one-form)
and a three form. The NS-R sectors contain  left-handed and right-handed
Majorana-Weyl gravitinos and spinors. This theory was shown to
coincide with the dimensionally reduced eleven-dimensional supergravity
action [6] not only for massless states [7] but for the massive ones as
well [1].
Up to date, the formulation of the eleven dimensional action is unique
and there is no known consistent modification of it. The alternative
formulation with a five-form although conjectured to exist by
 super five-branes in eleven dimensions is inconsistent as a field
theory [8].
What prevents carrying a duality transformation on the eleven-dimensional
action is the existence of a Chern-Simons term in the action involving
the three-form. It is not known how to apply a duality transformation
to an action where not only the field strength appear but the gauge
fields as well. Having mentioned that the  the type IIA supergravity
action is obtained by a simple dimensional reduction from the 
eleven-dimensional theory would seem to indicate that this form
of the theory is unique. This, however, is not the case as 
the reduction of the Chern-Simons form to ten dimensions 
could be manipulated in few interesting ways. The key observation
is that a three form in eleven-dimensiosn becomes a three-form and
a two-form in ten-dimensions. One can always write this term
in such a way that the action could be expressed in terms of the field
 strength of one of these forms (but
not both) thus allowing  a duality transformation on that field to
be performed. But this is not the whole story. The one-form
which is also present in the theory comes from the metric of the
eleven-dimensional theory. The surprising thing is that one can
find certain combination of fields such that the action could be
reexpressed in terms of two of the one, two or three forms. This
would allow for a family of duality transformations to be performed.
We shall show explicitely these transformations and that they
are consistent with supesymmetry.

Non-chiral $N=2$ supergravity in ten dimensions was obtained [7]
by trivialy reducing the eleven dimensional theory [6].
 The action is expressed in
terms of the bosonic fields $A_{\mu\nu}$ (or $A_2$),
$A_{\mu\nu\rho}$ (or $A_3$), $B_{\mu}$ (or $B$),
$\phi $ and the vielbein $e_{\mu}^a $.
Because of the presence of the one-form, two-form and three-form
we will denote this formulation by (1,2,3).
  The fermionic fields are
the gravitino $\psi_{\mu}$ and the spinor $\lambda $ both
of which are Majorana spinors. The action is given by [7]
$$
\eqalignno{
I &=\int d^{10}x e\Bigl( -{1\over {4 \kappa^2}} R(\omega (e) )
-{i\over 2} \overline{\psi}_{\mu} \Gamma^{\mu\nu\rho} D_{\nu} \psi_{\rho}
-{1\over {48}} e^{\kappa \phi} F_{\mu\nu\rho\sigma}^{\prime}
F^{\prime\mu\nu\rho\sigma} \cr
& \qquad +{1\over {12}} e^{-2\kappa \phi} F_{\mu\nu\rho}F^{\mu\nu\rho}
-{1\over 4} e^{3\kappa \phi} G_{\mu\nu}G^{\mu\nu} +{1\over 2}\partial_{\mu}
\phi \partial^{\mu}\phi \cr
& \qquad +{i\over 2} \overline{\lambda} \Gamma^{\mu} D_\mu \lambda
-{i\kappa \over {\sqrt 2}} \overline {\lambda} \Gamma^{11} \Gamma^{\mu}
\Gamma^{\nu}\psi_{\mu} \partial_{\nu}\phi \cr
&\qquad +{\kappa \over {8(12)^2}}e^{-1}\epsilon^{\mu_1 \cdots \mu_{10}}
F_{\mu_1 \cdots \mu_4}F_{\mu_5 \cdots \mu_8} A_{\mu_9\mu_{10}} \cr
& \qquad +{\kappa \over {96}} e^{\kappa \phi\over 2 } \bigl( 
\overline {\psi}_{\mu} \Gamma^{\mu\nu\alpha\beta\gamma\delta}\psi_{\nu}
+12 \overline{\psi}^{\alpha}\Gamma^{\beta\gamma}\psi^{\delta}
+{1\over \sqrt 2}\overline{\lambda} \Gamma^{\mu}\Gamma^{\alpha\beta
\gamma\delta}\psi_{\mu} +{3\over 4} \overline{\lambda} \Gamma^{\alpha
\beta\gamma\delta}\lambda \bigr) F_{\alpha\beta\gamma\delta}' \cr
&\qquad -{\kappa\over {24}}e^{-\kappa\phi\over 3}
\bigl( \overline{\psi}_{\mu}
\Gamma^{11} \Gamma^{\mu\nu\alpha\beta\gamma}\psi_{\nu} -6
\overline {\psi}^{\alpha} \Gamma^{11} \Gamma^{\beta}\psi^{\gamma}
-\sqrt 2 \overline {\lambda} \Gamma^{\mu} \Gamma^{\alpha\beta\gamma}
\psi_{\mu}\bigr) F_{\alpha\beta\gamma} \cr
&\qquad -{i\kappa\over 8} e^{3\kappa\phi\over 2}\bigl( 
\overline{\psi}_{\mu}\Gamma^{11} \Gamma^{\mu\nu\alpha\beta}\psi_{\nu}
+2\overline{\psi}^{\alpha}\Gamma^{11}\psi^{\beta} +{3\over \sqrt 2}
\overline{\lambda} \Gamma^{\mu}\Gamma^{\alpha\beta}\psi_{\mu}
+{5\over 4} \overline{\lambda} \Gamma^{11} \Gamma^{\alpha\beta}\lambda
\bigr) G_{\alpha\beta} \cr
&\qquad +{\rm quartic \ fermionic \ terms}\Bigr) & (1) \cr}
$$
where $G_{\mu\nu}$, $F_{\mu\nu\rho}$ and $F_{\mu\nu\rho\sigma}$
are field strengths of $B_{\mu}$, $A_{\mu\nu}$ and $A_{\mu\nu\rho}$
respectively. Because of the eleven dimensional origin of this
theory one has the modified field strength
$ F_{\mu\nu\rho\sigma}^{\prime} $
where
$$\eqalign{
G_{\mu\nu} &=2\partial_{[\mu}B_{\nu ]} \cr
F_{\mu\nu\rho} &=3\partial_{[\mu}A_{\nu\rho ]} \cr
F_{\mu\nu\rho\sigma}^{\prime} &=4\bigl( \partial_{[\mu}A_{\nu\rho\sigma ]}
+2B_{[\mu}F_{\nu\rho\sigma ]}\bigr) \cr}\eqno(2)
$$
As can be seen by compactifying the eleven-dimensional theory
working in a flat frame [5], we can write the field strength 
 $F'$ in terms
of a modified potential $A_3'$, where
$$\eqalign{
A_{\mu\nu\rho}^{\prime} &= A_{\mu\nu\rho}-6B_{[\mu}A_{\nu\rho ]} \cr
F_{\mu\nu\rho\sigma}^{\prime} &= 4 \bigl( \partial_{[\mu}A_{\nu\rho\sigma ]}'
+3 G_{[\mu\nu}A_{\rho\sigma ]} \bigr) \cr}\eqno(3)
$$
These identities will play a vital role in allowing for duality
transformations. The supersymmetry transformations are given by
$$\eqalign{
\delta e_{\mu}^a &= -i\overline{\epsilon} \Gamma^a \psi_{\mu} \cr
\delta \psi_{\mu} &= D_{\mu}(\omega ) -{1\over 32} e^{3\kappa\phi
\over 2}\bigl(\Gamma_{\mu}^{\ \nu\rho} -14 \delta_{\mu}^{\nu}\Gamma^{\rho}
\bigr) \Gamma^{11} \epsilon G_{\nu\rho} \cr
&\qquad +{i\over 48} e^{-\kappa\phi} \bigl( \Gamma_{\mu}^{\ \nu\rho\sigma} 
-9\delta_{\mu}^{\nu}\Gamma^{\rho\sigma}\bigr) \Gamma^{11} \epsilon
F_{\nu\rho\sigma} \cr
&\qquad +{i\over 128} e^{\kappa\phi\over 2} \bigl( \Gamma_{\mu}^
{\ \nu\rho\sigma\tau} -{20\over 3} \delta_{\mu}^{\nu} \Gamma^{\rho
\sigma\tau} \bigr) \epsilon F_{\nu\rho\sigma\tau}' +\cdots \cr
\delta B_{\mu} &= {i\over 2}e^{-{3\over 2}\kappa\phi}\bigl(
\overline{\psi}_{\mu} \Gamma^ {11} \epsilon -{\sqrt 2 \over 4}
\overline{\lambda} \Gamma_{\mu} \epsilon \bigr) \cr
\delta A_{\mu\nu} &=e^{\kappa\phi} \bigl( \overline{\psi}_{[\mu}
\Gamma_{\nu ]}\Gamma^{11} \epsilon -{1\over 2\sqrt 2} \overline{\lambda}
\Gamma_{\mu\nu} \epsilon \bigr) \cr
\delta A_{\mu\nu\rho} &= -{3\over 2}e^{-{\kappa\phi\over 2}}\bigl(
\overline{\psi}_{[\mu} \Gamma_{\nu\rho ]}\epsilon -{1\over
6\sqrt 2}\overline{\lambda} \Gamma^{11} \Gamma_{\mu\nu\rho}\epsilon
\bigr) \cr
&\qquad +6e^{\kappa\phi}
B_{[\mu}\bigl(\overline{\psi}_{\nu}\Gamma_{\rho ]}\Gamma^{11} \epsilon
-{1\over 2\sqrt 2} \overline{\lambda} \Gamma_{\nu\rho ]} \bigr) \cr
\delta \lambda &= {1\over \sqrt 2} D_{\mu}\phi (\Gamma^{\mu}
\Gamma^{11} \epsilon ) +{3\over 8\sqrt 2} e^{3\kappa\phi\over 2}
\Gamma^{\mu\nu}\epsilon G_{\mu\nu} \cr
&\qquad +{i\over 12\sqrt 2} e^{-\kappa\phi} \Gamma^{\mu\nu\rho}
\epsilon F_{\mu\nu\rho} +\cdots \cr
\delta \phi &= {i\over \sqrt 2} \overline{\lambda} \Gamma^{11} \epsilon
\cr}\eqno(4)
$$ 

Another important piece is the Chern-Simons term which can be
written in terms of differential forms as $ \int A_2 \wedge 
dA_3 \wedge dA_3 $ where
$A_2$ and $A_3$ stand for the two- and three-forms:
$A_2=A_{\mu\nu}dx^{\mu}\wedge dx^{\nu}$, and 
$A_3=A_{\mu\nu\rho}dx^{\mu}\wedge dx^{\nu}\wedge dx^{\rho}$. This 
can be reexpressed in such a way that  
 $A_{\mu\nu}$ appears only through  its field strength $F_{\mu\nu\rho}$.
We derive this by using
$$
A_2 \wedge dA_3 \wedge dA_3 =d(A_2\wedge A_3\wedge dA_3)
 -dA_2 \wedge A_3 \wedge dA_3 \eqno(5)
$$
and discarding the surface term after integration.    
Next, although the field $B_{\mu}$ does not appear in the
Chern-Simons term, it appears explicitely in the
field strength $F_{\mu\nu\rho\sigma}'$ in eq (2). If  
equation (3) is used instead of (2), then $B_{\mu}$ appears
only through its field strength $G_{\mu\nu}$ but then the 
Chern-Simons term must  be expressed in terms of
$A_{\mu\nu\rho}'$. It is not difficult to show that
$$\eqalign{
A_2 \wedge dA_3 \wedge dA_3 &= A_2 \wedge d A_3' \wedge dA_3'
 +6 A_2\wedge A_2 \wedge dB \wedge dA_3' \cr
&\qquad +12 A_2\wedge A_2\wedge A_2 \wedge dB \wedge dB \cr 
&\qquad +6 d\bigl(A_2\wedge A_2\wedge B \wedge (dA_3' +4A_2\wedge dB )\bigr)
 \cr} \eqno(6)
$$
Discarding the surface term, we see that the action (1) is
expressible in terms of $A_2$, $dA_3'$ and $dB$. From all of 
these considerations it is very suggestive that we can apply
 duality transformations to the following fields $(A_6, A_2)$,
 or $(A_3, A_2)$ or $(B, A_7)$. We now consider these
transformations one at a time.

To obtain the dual theory where the two-form  is replaced
with a six-form, we add to the action (1) the term
$$
{1\over {3!6!}}\int A_6 \wedge dF_3  \eqno(7)
$$
where $A_6 = A_{\mu_1\cdots \mu_6}dx^{\mu_1}\wedge \cdots \wedge
dx^{\mu_6} $
is a six-form and $F_3$ is a three-form, $F_3 =F_{\mu\nu\rho}
dx^{\mu}\wedge dx^{\nu}\wedge dx^{\rho} $, 
which in (1) 
is not assumed now to be a field strength. The equation  
of motion of $A_6$ forces  $F_3$, locally, to be $dA_2$.
Integrating by parts and discarding the surface term, 
eq (7) can be rewritten in the form
$$
{1\over {3!6!}} \int F_3 \wedge dA_6   \eqno (8)
$$
Since $F_3$ appears in the action (1) and (8) at most
quadratically, we can perform the $F_3$ gaussian integration to
obtain the dual version as a function of $A_6$. 
Therefore, the action in the form (1) plus (7) can give either one
of the two dual actions, depending on what is integrated
first, $A_6$ or $F_3$.  The  
supersymmetry transformations of  the combined action can be found
as follows [9]. The supersymmetry transformations of $F_3$ are
taken to be identical to those of $d\delta A_2 $ as given
in eq (2) (without identifying $F_3$ with $dA_2$ ), then
the action (1) will be invariant except for one term proportional
to $dF_3$ which does not vanish now because the Bianchi 
identity is no  longer available. The non-invariant term will be cancelled
by the transformation of the new term (7) which is also
proportional to $\int \delta A_6 \wedge dF_3 $ . This determines 
$ \delta A_6 $ to be given by
$$
\delta A_{\mu_1 \cdots \mu_6} =-3i e^{-\kappa \phi }
\bigl( \overline {\epsilon} \Gamma_{[\mu_1 \cdots \mu_5}\psi_{\mu_6]}
-{i\over {6\sqrt 2}} \overline{\epsilon} \Gamma_{\mu_1 \cdots \mu_6}
\Gamma^{11} \lambda \bigr) \eqno(9)
$$
and explicitely shows that the action (1) plus (7) admits a
duality transformation between the two-form and the six-form. 
The duality transformation is at the level of the action and not
only the equations of motion. As
the field $F_{\mu\nu\rho}$ appears at most quadratically, doing
the gaussian integration for $F_{\mu\nu\rho}$, or solving its equation
of motion and substituting back into the action, are equivalent.
The equation of motion gives
$$
M_{\alpha\beta\gamma}^{\mu\nu\rho}F_{\mu\nu\rho}=X_{\alpha\beta\gamma}
\eqno(10)
$$
where the tensors $M_{\alpha\beta\gamma}^{\mu\nu\rho}$ and 
$X_{\alpha\beta\gamma}$ are given by
$$\eqalignno{
M_{\alpha\beta\gamma}^{\mu\nu\rho} &= \Bigl( {1\over {3!6}} 
e^{-2\kappa \phi}(1-e^{3\kappa\phi}B_{\sigma}B^{\sigma})
\delta_{\alpha\beta\gamma}^{\mu\nu\rho} +{1\over 4}e^{\kappa \phi}
B_{[\alpha}\delta_{\beta\gamma}^{[\nu\rho}B^{\mu ]} \Bigr) & (11)\cr}
$$
$$\eqalignno{
X_{\alpha\beta\gamma} &=-{1\over {216}} \epsilon_{\alpha\beta\gamma}^
{\ \ \ \ \ \mu_1 \cdots \mu_7} \bigl( {1\over {7!}} F_{\mu_1\cdots \mu_7}
+A_{\mu_1\mu_2\mu_3}\partial_{\mu_4}A_{\mu_5\mu_6\mu_7} \bigr) \cr
&\qquad +{\kappa \over {24}} e^{-{1\over 3}\kappa \phi }\bigl(
\overline{\psi}_{\mu} \Gamma^{11} \Gamma^{\mu\nu}_{\ \ \alpha\beta\gamma}
\psi_{\nu}
-6\overline{\psi}_{[\alpha}\Gamma^{11}\Gamma_{\beta}\psi_{\gamma ]}
-\sqrt 2 \overline{\lambda} \Gamma^{\mu}\Gamma_{\alpha\beta\gamma}
\psi_{\mu}\bigr) \cr
&\qquad -{\kappa \over {12}}e^{{1\over 2}\kappa\phi }\bigl(
\overline{\psi}_{\mu}\Gamma^{\mu\nu\rho}_{\ \ \  \alpha\beta\gamma} \psi_{\nu}
+12\overline{\psi}^{\rho}\Gamma_{[\alpha \beta}\psi_{\gamma ]} \cr
&\qquad \qquad \qquad +{1\over {\sqrt 2}} \overline{\lambda} \Gamma^{\mu}
\Gamma^{\rho}_{\ \alpha\beta\gamma}\psi_{\mu} +{3\over 4} \overline{\lambda}
\Gamma^{\rho}_{\ \alpha\beta\gamma}\lambda \bigr) B_{\rho} &(12)\cr} 
$$
and we have denoted $F_{\mu_1\cdots\mu_7}=7\partial_{[\mu_1}A_{\mu_2
\cdots\mu_7 ]} $. 

Solving equation (10) for $F_{\mu\nu\rho}$ gives
$$
F_{\mu\nu\rho}=M_{\ \ \ \mu\nu\rho}^{-1\alpha\beta\gamma}X_{\alpha\beta\gamma}
\eqno(13)
$$
where the tensor $ M_{\ \ \ \mu\nu\rho}^{-1\alpha\beta\gamma}$ is the inverse
of $M_{\alpha\beta\gamma}^{\mu\nu\rho}$:
$$M_{\ \ \ \mu\nu\rho}^{-1\alpha\beta\gamma}M_{\alpha\beta\gamma}^{\kappa
\lambda\sigma} ={1\over {3!}}\delta_{\mu\nu\rho}^{\kappa\lambda\sigma}
\eqno(14)
$$
The explicit form of $M^{-1}$ is
$$
M_{\ \ \ \mu\nu\rho}^{-1 \alpha\beta\gamma}={6
e^{2\kappa\phi}\over {1-e^{3\kappa\phi}B_{\sigma}B^{\sigma}}}\Bigl( 
{1\over {3!}}\delta_{\alpha\beta\gamma}^{\mu\nu\rho}-{3\over
2}e^{\kappa \phi} B_{[\mu} \delta_{\nu\rho
]}^{[\alpha\beta}B^{\gamma ]}\Bigr) \eqno(15)
$$
Therefore to obtain the dual action from (1) plus (7), we discard
all the $F_{\mu\nu\rho}$ contributions and replace them with
$$
-{1\over 2} X_{\alpha\beta\gamma} M_{\ \ \ \mu\nu\rho}^{-1 \alpha\beta\gamma}
X_{\mu\nu\rho}   \eqno(16)
$$
The action in (16) is a non-polynomial function of $B_{\mu}$.
It is an interesting question to find whether some field
redefinitions involving the dilaton can change the dependence
to a polynomial one.

To find the $N=2$ supergravity action where the three-form
is replaced with a five-form we proceed as before. First,
we write the action (1) in such a way that  the three-form
appears only through its field strength. We 
use eq (3) for $F_{\mu\nu\rho\sigma}'$, and
write it as $F_{\mu\nu\rho\sigma}+12 G_{[\mu\nu}A_{\rho\sigma ]}$.
Then we assume that 
$F_{\mu\nu\rho\sigma }$ is an independent field
and not the field strength of $A_{\mu\nu\rho}'$, and
add the following term to the action:
$$
{1 \over {4!5!}}\int A_5 \wedge dF_4  \eqno(17)
$$
where $A_5=A_{\mu_1\cdots \mu_5}dx^{{\mu}_1}\wedge \cdots 
\wedge dx^{{\mu}_5} $. The
$A_5$ equation implies, locally, that $F_4=dA_3' $ and this gives again
the action (1). If, however, we integrate eq (17) by parts, 
and then do the gaussian integration of $F_{\mu\nu\rho\sigma}$
we will be left with an action in terms of the dual field
strength $F_{\mu_1\cdots\mu_6}$. To restore the supersymmetry
invariance after adding (17) to the action (1) we assume
that $\delta F_{\mu\nu\rho\sigma} =4\partial_{[\mu} \delta
A_{\nu\rho\sigma ]}$, then the extra terms that spoil the invariance
of the action (1) are cancelled by those arising from the
non-invariance of the term (17). This
is achieved by taking
$$
\delta A_{\mu_1\cdots\mu_5}={5\over 2}i e^{{1\over 2}\kappa \phi }
\overline{\epsilon} \Gamma^{11}
\Gamma_{[\mu_1\cdots\mu_4}\psi_{\mu_5 ]} \eqno(18)
$$
The sum of the actions (1) and (17)  gives both
dual actions depending on the order of integration and is 
invariant under the new supersymmetry transformations.

The gaussian integration of $F_{\mu\nu\rho\sigma}$ gives
$$
{1\over 2}  X_{\mu\nu\rho\sigma}M_{\ \ \ \alpha\beta\gamma\delta}^{-1
\mu\nu\rho\sigma} X^{\alpha\beta\gamma\delta} \eqno(19)
$$
where $X_{\mu\nu\rho\sigma}$ is defined by
$$\eqalign{
X_{\mu\nu\rho\sigma} &= {\kappa \over {4!}} \epsilon_{\mu\nu\rho\sigma}
^{\ \ \ \ \ \mu_1\cdots\mu_6}\bigl( {1\over 6!} F_{\mu_1\cdots\mu_6}
+{1\over 16} A_{\mu_1\mu_2}A_{\mu_3\mu_4}G_{\mu_5\mu_6} \bigr) \cr
&\qquad +{\kappa \over 96}e^{{1\over 2}\kappa \phi }\bigl( \overline
{\psi}_{\rho}\Gamma^{\alpha\beta}_{\ \ \mu\nu\rho\sigma}\psi_{\beta}
+12 \overline{\psi}_{[\mu} \Gamma_{\nu\rho}\psi_{\sigma ]}
+{1\over {\sqrt 2}} \overline{\lambda}\Gamma^{\alpha}\Gamma_{
\mu\nu\rho\sigma}\psi_{\alpha} +{3\over 4} \overline{\lambda}
\Gamma_{\mu\nu\rho\sigma}\lambda\bigr) \cr
&\qquad -{1\over 2} e^{\kappa\phi} G_{[\mu\nu}A_{\rho\sigma ]}\cr}\eqno(20)
$$
and the matrix $M^{-1}$ is the inverse of 
$$
M_{\mu\nu\rho\sigma}^{\alpha\beta\gamma\delta}=({1\over 4!})^2
\Bigl( e^{\kappa\phi}\delta_{\mu\nu\rho\sigma}^{\alpha\beta\gamma
\delta} -\kappa \epsilon_{\mu\nu\rho\sigma}^{\ \ \ \ \alpha
\beta\gamma\delta\lambda\tau}A_{\lambda\tau} \Bigr) \eqno(21)
$$
defined by
$$
M_{\ \ \ \mu\nu\rho\sigma}^{-1\alpha\beta\gamma\delta}
M_{\alpha\beta\gamma\delta}^{\kappa\lambda\tau\eta}
={1\over {4!}}\delta_{\mu\nu\rho\sigma}^{\kappa\lambda\tau\eta}
\eqno(22)
$$
The explicit expression of $M^{-1}$ is too long to give here.
The field strength $F_6$ is given by 
$$
F_{\mu_1\cdots\mu_6}=6\partial_{[\mu_1} A_{\mu_2\cdots\mu_6 ]} \eqno(23)
$$  

Therefore, to obtain the dual action we discard all the terms
containing $F_{\mu\nu\rho\sigma}$ and replace them with  (19).
This completes the derivation of the dual action where the
three-form is replaced by the five-form.

By writing the $N=2$ supergravity action IIA in such a way that
the one-form $B$ appears only through its field strength
required a redefinition of the three-form. The procedure
of obtaining the action where the one-form is replaced with
the seven-form is the same as before. We first manipulate
the action (1) so that the field $B_{\mu}$ appears only
through its field strength $G_{\mu\nu}$ then we assume that
$G_{\mu\nu}$ is an independent field and add a term to the
action (1) of the form:
$$
{1\over{2!7!}} \int A_7 \wedge d G  \eqno(24)
$$
where we have defined the seven-form 
$A_7=A_{\mu_1\cdots\mu_7}dx^{\mu_1}\wedge \cdots \wedge dx^{\mu_7}$
Integrating the $A_7$ field out implies the constraint
$dG=0$  whose solution , locally, is
$G_{\mu\nu}=2\partial_{[\mu}B_{\nu ]}$ and this takes us back
to the action (1). Integrating the action (24) by parts
and discarding the surface term we obtain
$$
{{1\over {2!7!}}} \int dA_7 \wedge G  \eqno(25)
$$
The field $F_{\mu\nu\rho\sigma}'$ in the action (1) is
taken to be of the form (3) and the Chern-Simons term is
rearranged to be given by (5). Then the full action is
at most quadratic in $G_{\mu\nu}$ and the gaussian
integration can be performed. This will give the dual
action expressed in terms of the field strength of $A_7$.
The non-invariance of (1) under the supersymmetry transformations
due to the removal of the identificaiton $G=dB$ is cancelled 
by the varriation of (24) provided one identifies the
varriation of $G$ with
$$
\delta G_{\mu\nu}=2\partial_{[\mu}\delta B_{\nu ]}  \eqno(26)
$$
and the varriation of $A_7$ with
$$\delta A_{\mu_1\cdots\mu_7} =e^{{3\over 2}\kappa \phi}\bigl( 
-{7\over 2} \overline{\epsilon}\Gamma_{[\mu_1\cdots\mu_6}\psi_{\mu_7 ]}
+{\sqrt 2\over 8} \overline{\epsilon} \Gamma_{\mu_1\cdots\mu_7}
\Gamma^{11}\lambda \bigr) \eqno(27)
$$
The gaussian integration of $G_{\mu\nu}$ gives
$$
-{1\over 4} X_{\mu\nu}M_{\ \ \ \alpha\beta}^{-1\mu\nu}X^{\alpha\beta}
\eqno(28)
$$
where $X_{\mu\nu}$ is defined by
$$\eqalign{  
X_{\mu\nu} &=-{1\over 8}e^{\kappa\phi}A^{\rho\sigma}\bigl(
4\partial_{[\mu} A_{\nu\rho\sigma ]}'\bigr) \cr
&\qquad +{3\kappa\over {16}} \bigl( \overline{\psi}_{\alpha}
\Gamma^{\alpha\beta}_{\ \ \mu\nu\rho\sigma}\psi_{\beta} 
+12\overline{\psi}_{[\mu}\Gamma_{\nu\rho}\psi_{\sigma ]} 
+{1\over {\sqrt 2}}\overline{\lambda} \Gamma^{\alpha}
\Gamma_{\mu\nu\rho\sigma}\psi_{\alpha} +{3\over 4}
\overline{\lambda}\Gamma_{\mu\nu\rho\sigma}\lambda \bigr) 
A^{\rho\sigma} \cr
&\qquad -{i\kappa \over 8}e^{{3\over 2}\kappa \phi}\bigl(
\overline{\psi}_{\alpha}\Gamma^{11}\Gamma^{\alpha\beta}_{\ \ \mu\nu}
\psi_{\beta} +2\overline{\psi}_{[\mu}\Gamma^{11} \psi_{\nu ]}
+{3\over 2} \overline{\lambda} \Gamma^{\alpha}\Gamma_{\mu\nu}\psi_{\alpha}
+{5\over 4}\overline{\lambda} \Gamma^{11}\Gamma_{\mu\nu}\lambda \bigr)\cr
&\qquad +\epsilon_{\mu\nu}^{\ \ \ \mu_1\cdots\mu_8} \Bigl(
{1\over{2!7!}} \partial_{\mu_1}A_{\mu_2\cdots\mu_8}
-{\kappa \over 192}A_{\mu_1\mu_2}\cdots A_{\mu_7\mu_8} \Bigr) \cr}\eqno(29)
$$
and the tensor $M_{\ \ \ \mu\nu}^{-1\alpha\beta}$ is the inverse
of 
$$\eqalign{
M_{\mu\nu}^{\alpha\beta} &= {1\over 4}e^{3\kappa \phi}\bigl(
1+{3\over
2}A_{\rho\sigma}A^{\rho\sigma}\bigr)\delta_{\mu\nu}^{\alpha\beta}
+A_{\mu\nu}A^{\alpha\beta} \cr
&\qquad -4\delta_{[\mu}^{[\alpha}A_{\nu ]\rho}A^{\beta ]\rho}
-{\kappa\over 96}\epsilon_{\ \ \ \mu\nu\mu_1\cdots\mu_6}^{\alpha
\beta}A_{\mu_1\mu_2}A_{\mu_3\mu_4}A_{\mu_5\mu_6} \cr}\eqno(30)
$$
The inverse of $M$ is defined by:
$$
M_{\ \ \ \mu\nu}^{-1\alpha\beta}M_{\alpha\beta}^{\rho\sigma}={1\over {2!}}
\delta_{\mu\nu}^{\rho\sigma} \eqno(31)
$$  
but again the explicit expression is too long to give here.
Finally, $G_{\mu\nu}$ is related to its dual by the relation
$$
G_{\mu\nu}=M_{\ \ \ \mu\nu}^{-1\alpha\beta}X_{\alpha\beta}\eqno(32)
$$
The dual action is obtained by discarding all the $G_{\mu\nu}$ 
contributions in (1) plus (25) and replacing them with (28).
This completes the derivation of the dual action where the one-form
is replaced with a seven-form.

Therefore, we have  shown that the original formulation
of $N=2$ supergravity type IIA given in terms of a one-form, a
two-form
and a three-form  (we denote this by (1,2,3)), admits three
other dual formulations. In the first, the two-form is replaced
with a six-form giving rise to a formulation in terms of a one-form,
a six-form and a three-form (denoted by (1,6,3)). In the second the 
three-form is replaced with a five-form giving rise to 
a formulation in terms of (1,2,5) forms. Finally, in the
third the one-form is replaced with a seven-form
giving rise to the (7,2,3) formulation. 
It is easy to see that the (1,2,5) formulation depends on
the three-form through its field strength suggesting that
it is possible to find a duality transformation  that takes the
one-form
to a sevem-form. This will give  the (7,2,5) formulation.
 This can also be reached by performing a duality transformation
on the three-form in the (7,2,3) formulation 
as it appears only through its field strength.
This also implies that the (7,2,5) formulation can be
reached by applying a double duality transformation to
the one-form and three-form simultaneously. If we arrange
the (1,2,3), (7,2,3), (7,2,5) and (1,2,5) formulations
at the corners of a square in a clockwise fashion, then 
all adjacent vertices could be transformed to each other
by a simple duality transformation, and the opposite edges by
a double duality transformation. But it seems that the (1,6,3)
formulation can only be connected to the (1,2,3) formulation
as it depends on the one-form and three-form explicitely.

The fact that only certain combinations of field configurations
are allowed seems to indicate some consistency conditions.
It will be useful to deduce such conditions as some projections
on physical states of the spectrum of the Dirichlet-branes [3]. 
It will also be useful to find out whether such conditions
give only the theories mentioned here, or whether they allow
for other combinations signalling the possibility of new
theories in ten-dimensions and may be in eleven. 
\vskip.4truecm
{\bf \noindent References}
\vskip.2truecm

\item{[1]}C. M. Hull and P. K. Townsend, {\sl Nucl. Phys.}
{\bf B438} (1995) 109;\br
E. Witten, {\sl Nucl. Phys.} {\bf B 443} (1995) 85.

 \item{[2]}M. J. Duff and X. J. Lu, {\sl Nucl. Phys.} 
{\bf B354} (1991) 141; {\sl Nucl. Phys.} {\bf B390}
(1993) 276;\br
M. J. Duff, R. R. Khuri and X. J. Lu, {\sl Phys. Rept.}
{\bf 259} (1995) 213.

\item{[3]}J. Polchinski, {\sl Dirichlet-Branes and Ramond-Ramond
Charges} hepth/9510017.

\item{[4]} A. H. Chamseddine, {\sl Nucl.Phys.} {\bf B185 } (1981)
403;\br
E. A. Bergshoeff, M. de Roo, B. de Witt and P. van Niewenhuizen,
{\sl Nucl. Phys.} {\bf B195} (1982) 97;\br
G. F. Chapline and N. S. Manton, {\sl Phys. Lett} {\bf 120B} (1983)
105.

\item{[5]}A. H. Chamseddine, {\sl Phys. Rev.} {\bf D24} (1981)
3065;\br
S. Gates and H. Nishino, {\sl Phys. Lett.} {\bf 173B} (1986)
46 and 52;\br
A. H. Chamseddine and P. Nath, {\sl Phys. Rev.}
{\bf D34} (1986) 3769;\br
E. A. Bergshoef and M. de Roo, {\sl Nucl. Phys.} {\bf B328}
(1989) 439.

\item{[6]}E. Cremmer, B. Julia and J. Scherk, {\sl Phys. Lett }
{\bf 76B} (1978) 409. 

\item{[7]} I. C. Campbell and P. West, {\sl Nucl. Phys.} 
{\bf B243} (1984) 112.

\item{[8]} H. Nicolai, P. K. Townsend and P. van Niewenhuizen,
{\sl Lett. Nuovo. Cimento} {\bf 30} (1980) 179.

\item{[9]} H. Nicolai and P. K. Townsend, {\sl Phys. Lett.}
{\bf 98B} (1981) 257.

\end